\documentclass[manuscript]{aastex}

\slugcomment{Not to appear in Nonlearned J., 45.}

\shorttitle{}
\shortauthors{Fukui et al.}

\usepackage{color}
\begin{document}


\title{High-mass star formation triggered by collision between CO filaments in N159 West in the Large Magellanic Cloud}


\author{Yasuo Fukui\altaffilmark{1}, Ryohei Harada\altaffilmark{2}, Kazuki Tokuda\altaffilmark{2}, Yuuki Morioka\altaffilmark{2}, Toshikazu Onishi\altaffilmark{2}, Kazufumi Torii\altaffilmark{1}, Akio Ohama\altaffilmark{1},  Yusuke Hattori\altaffilmark{1}, Omnarayani Nayak\altaffilmark{3}, Margaret Meixner\altaffilmark{4,3}, Marta Sewi\l{o}\altaffilmark{5,3}, Remy Indebetouw\altaffilmark{6,7}, Akiko Kawamura\altaffilmark{8}, Kazuya Saigo\altaffilmark{8}, Hiroaki Yamamoto\altaffilmark{1}, Kengo Tachihara\altaffilmark{1}, Tetsuhiro Minamidani\altaffilmark{10}, Tsuyoshi Inoue\altaffilmark{9}, Suzanna Madden\altaffilmark{11}, Maud Galametz\altaffilmark{12,13}, Vianney Lebouteiller\altaffilmark{11}, Norikazu Mizuno\altaffilmark{14,6}, and C.-H. Rosie Chen\altaffilmark{15}}

\email{}


\altaffiltext{1}{Department of Physics, Nagoya University, Chikusa-ku, Nagoya 464-8602, Japan}
\altaffiltext{2}{Department of Physical Science, Graduate School of Science, Osaka Prefecture University, 1-1 Gakuen-cho, Naka-ku, Sakai, Osaka 599-8531, Japan}
\altaffiltext{3}{The Johns Hopkins University, Department of Physics and Astronomy, 366 Bloomberg Center, 3400 N. Charles Street, Baltimore, MD 21218, USA}
\altaffiltext{4}{Space Telescope Science Institute, 3700 San Martin Drive, Baltimore, MD 21218, USA}
\altaffiltext{5}{Space Science Institute, 4750 Walnut Street, Suite 205, Boulder, CO 80301, USA}
\altaffiltext{6}{Department of Astronomy, University of Virginia, PO Box 400325, Charlottesville, VA, 22904, USA}
\altaffiltext{7}{National Radio Astronomy Observatory, 520 Edgemont Rd, Charlottesville, VA, 22903, USA}
\altaffiltext{8}{National Astronomical Observatory of Japan, Mitaka, Tokyo 181-8588, Japan}
\altaffiltext{9}{Division of Theoretical Astronomy, National Astronomical Observatory, Japan}
\altaffiltext{10}{Nobeyama Radio Observatory, 462-2 Nobeyama Minamimaki-mura, Minamisaku-gun, Nagano 384-1305, Japan}
\altaffiltext{11}{Laboratoire AIM, CEA, Universite Paris VII, IRFU/Service d'Astrophysique, Bat. 709, 91191 Gif-sur-Yvette, France}
\altaffiltext{12}{Institute of Astronomy, University of Cambridge, Madingley Road, Cambridge CB3 0HA, UK}
\altaffiltext{13}{European Southern Observatory, Karl-Schwarzschild-Str. 2, D-85748 Garching-bei-M\"{u}nchen, Germany}
\altaffiltext{14}{Department of Astronomy, School of Science, The University of Tokyo, 7-3-1 Hongo, Bunkyo-ku, Tokyo 133-0033, Japan}
\altaffiltext{15}{Max Planck Institute for Radio Astronomy, Auf dem Huegel 69, Bonn 53121, Germany}


\begin{abstract}
We have carried out $^{13}$CO($J$=2--1) observations of the active star-forming region N159 West in the LMC with ALMA. We have found that the CO distribution at a sub-pc scale is highly elongated with a small width. These elongated clouds called ``filaments'' show straight or curved distributions with a typical width of 0.5--1.0\,pc and a length of 5--10\,pc. All the known infrared YSOs are located toward the filaments. We have found broad CO wings of two molecular outflows toward young high-mass stars in N159W-N and N159W-S, whose dynamical timescale is $\sim$10$^{4}$ yrs. This is the first discovery of protostellar outflow in external galaxies. For N159W-S which is located toward an intersection of two filaments we set up a hypothesis that the two filaments collided with each other $\sim$10$^{5}$ yrs ago and triggered formation of the high-mass star having $\sim$37 $M_{\odot}$. The colliding clouds show significant enhancement in linewidth in the intersection, suggesting excitation of turbulence in the shocked interface layer between them as is consistent with the magneto-hydro-dynamical numerical simulations \citep{Inoue2013}. This turbulence increases the mass accretion rate to $\sim6\times10^{-4}$ $M_{\odot}$ yr$^{-1}$, which is required to overcome the stellar feedback to form the high-mass star.
\end{abstract}


\keywords{ISM: clouds --- ISM: molecules --- ISM: kinematics and dynamics --- stars: formation}

\section{Introduction \label{s.intro}}

High-mass stars are influential in galactic evolution by dynamically affecting and ionizing the interstellar medium, and also by chemically enriching heavy elements via supernova explosions. It is of fundamental importance to understand the physical processes in the evolution of molecular clouds where high-mass stars are forming. There have been numerous works on high-mass star formation in the literature \citep[for reviews see e.g.,][]{Zinnecker2007,Tan2014}. In spite of these works we have not yet understood how high-mass star formation takes place. One of the promising candidates where young high-mass stars are forming is the very dense and massive cores such as infrared dark clouds in the Milky Way \citep{Peretto2013}. Another possible candidate is the compressed layer formed in cloud-cloud collisions. Observations of a few super star clusters and smaller H$\,${\sc ii} regions in the Milky Way have shown signs of triggered formation of high-mass stars in the collision-compressed layers \citep[e.g.,][]{Furukawa2009,Torii2011,Torii2015,Fukui2014,Fukui2015}. Magneto-hydro-dynamical numerical simulations of two colliding molecular clouds by \citet{Inoue2013} have shown that turbulence is excited and the magnetic field is amplified in the collision-shocked layer between the clouds. These turbulence and magnetic field increase the mass accretion rate, favoring high-mass star formation.

The difficulty in studying young high-mass stars lies in the considerably small number of young high-mass stars as compared with low-mass stars in the solar vicinity; this is in part due to the lower frequency of high-mass stars and the heavy sightline contamination in the Galactic disk. ALMA is now opening a new possibility to explore high-mass star formation in external galaxies by its unprecedented sensitivity and resolution, having a potential to revolutionize our view of high-mass star formation. The Large and Small Magellanic Clouds (LMC and SMC), at a distance of 50\,kpc \citep{Schaefer2008} and 61\,kpc \citep{Szewczyk2009}, are actively forming high-mass stars. The LMC is an ideal laboratory to see the evolution of stars and clouds thanks to the non-obscured face-on view \citep{Subramanian2010} of all the GMCs in a single galaxy \citep[for a review][]{Fukui2010}. A $^{12}$CO($J$=1--0) survey for GMCs with NANTEN 4m telescope \citep{Fukui1999,Mizuno2001,Yamaguchi2001,Fukui2008} provided a sample of nearly 300 GMCs at 40\,pc resolution and led to an evolutionary scheme from starless GMCs (Type I) to active star-forming GMCs (Type III) over a timescale of 20 Myrs \citep{Fukui1999,Kawamura2009}. Aiming at revealing the finer-scale details of the molecular gas in the LMC, we have commenced systematic CO observations by using ALMA at sub-pc resolution.

Among the nearly 300 GMCs over the LMC obtained with NANTEN, N159 is the brightest one with H$\,${\sc ii} regions. Infrared studies have revealed nearly twenty young high-mass stars in N159 with Spitzer and Herschel (\citealt{Chen2010}, \citealt{Wong2011}, \citealt{Carlson2012} and references therein; \citealt{Seale2014}), where two clumps N159 East and West are active in star formation. \citet{Mizuno2010} showed that the CO $J$=4--3/$J$=1--0 ratio shows enhancement toward the molecular peak without a well-developed H$\,${\sc ii} region in N159 West (N159W). This high excitation condition suggests that N159W is possibly on the verge of high-mass star formation, and thus the initial condition of high-mass star formation may still hold. The preceding observations with Australia Telescope Array (ATCA), while low in resolution (HPBW$\sim6\arcsec$), presented some hint of small-scale clumps and filaments in N159W\citep{Seale2012}. N159W is therefore the most suitable target for the purpose of witnessing the oneset of high-mass star formation. 

We present the first results of the ALMA observations of N159W in this Letter mainly based on the $^{13}$CO($J$=2--1) data.

\section{Observations \label{s.obs}}

We carried out ALMA Cycle 1 Band 3 (86--116\,GHz) and Band 6 (211--275\,GHz) observations toward N159W both with the main array 12m antennas and the Atacama Compact Array (ACA) 7m antennas. The observations centered at ($\alpha_{J2000.0}$, $\delta_{J2000.0}$) = (5$^{\rm h}$39$^{\rm m}$35\fs34, -69\arcdeg45\arcmin33\farcs2), were carried out between October 2013 to May 2014. The target molecular lines were $^{13}$CO($J$=1--0), C$^{18}$O($J$=1--0), CS($J$=2--1), $^{12}$CO($J$=2--1), $^{13}$CO($J$=2--1) and C$^{18}$O($J$=2--1) with a bandwidth of 58.6 MHz (15.3 kHz $\times$ 3840 channels). We used a spectral window for the observations of the continuum emission among the four with a bandwidth of 1875.0 MHz (488.3 kHz $\times$ 3840 channels). The radio recombination lines of H30$\alpha$ and H40$\alpha$ were also included in the windows. The projected baseline length of the 12m array ranges from 16\,m to 395\,m. The ACA covers 9\,m to 37\,m baselines. The calibration of the complex gains was carried out through observations of seven quasars, phase calibration on four quasars, and flux calibration on five solar system objects. For the flux calibration of the solar system objects, we used the Butler-JPL-Horizons 2012 model (https://science.nrao.edu{\slash}facilities{\slash}alma{\slash}aboutALMA{\slash}Technology{\slash}ALMA\_Memo\_Series{\slash}alma594{\slash}abs594). The data were reduced using the Common Astronomy Software Application (CASA) package (http://casa.nrao.edu), and visibility imaged. 
We used the natural weighting for both the Band 3 and Band 6 data, providing synthesized beam sizes of $\sim$2\farcs5 $\times$ 1\farcs8 (0.6 pc $\times$ 0.4 pc at 50kpc) and $\sim$1\farcs3 $\times$ 0\farcs8 (0.3 $\times$ 0.2 pc), respectively. 
The rms noises of molecular lines of Band 3 and band 6 are $\sim$40 mJy beam$^{-1} $and $\sim$20 mJy beam$^{-1}$, respectively, in emission-free channels.
The comparison of the cloud mass derived from the ALMA observation with that of a single dish observation described in Section \ref{s.r.filaments} suggests that the missing flux of the present ALMA observation is not significant.

\section{Results \label{results}}
\subsection{A complex filamentary structure \label{s.r.filaments}}

Figure \ref{fig1} shows the $^{13}$CO velocity integrated intensity image of the $J$=2--1 transition.  
The distribution of the $^{13}$CO emission is highly filamentary. 
The filaments, having often straight or curved distribution, have a typical length of 5--10\,pc and a width of 0.5--1.0\,pc defined as a full-width of the emission area at the 3$\sigma$ level of the intensity integrated over a range of 234 to 240\,km\,s$^{-1}$, which may be analogous to the dominance of filaments in the interstellar medium of the solar vicinity \citep[e.g.,][]{Andre2013,Molinari2010}, possibly suggesting that filaments are ubiquitous in other galaxies as well. 
More details of the filaments will be published separately.
The most active star formation is found in two regions as denoted by N159W-N and N159W-S in Figure \ref{fig1}, both of which are associated with enhanced $^{13}$CO emission. 

The cloud mass is estimated from the $^{12}$CO($J$=2--1) intensity by assuming a conversion factor from the  $^{12}$CO($J$=1--0) intensity to the column density of X(CO)=$7\times10^{20}$cm$^{-2}$ \citep{Fukui2008} and the typical $^{12}$CO($J$=2--1)/$^{12}$CO($J$=1--0) ratio toward H{\sc ii} regions of 0.85 (the ratio in the Orion-KL region of \citealt{Nishimura2015}).
 We also assumed the absorption coefficient per unit dust mass at 1.2\,mm and the dust-to-gas mass ratio to be 0.77\,cm$^2$\,g$^{-1}$ and $3.5\times10^{-3}$, respectively to derive the gas mass from the dust emission (Herrera et al. 2013).
In total, the filaments have molecular mass of 2.4 $\times$ 10$^{5}$ $M_{\odot}$ in N159W corresponding to 35 \% of the total mass which is estimated by the lower resolution study \citep{Minamidani2008,Mizuno2010}. 
We define the N159W-N and N159W-S clumps at the 5\,$\sigma$ level of Band 6 continuum (white contours in Figure 1), and the masses of these clumps are estimated to be $2.9\times10^4$\,$M_\odot$ and $4.1\times10^3$\,$M_\odot$, respectively, by assuming the dust temperature 20\,K. Their masses derived from the CO emission are $1.5\times10^4$\,$M_\odot$ and $4.2\times10^3$\,$M_\odot$, respectively, as is consistent with the dust-emission estimate.

\subsection{Outflows \label{s.r.outflows}}

We have discovered two molecular outflows having velocity span of 10--20\,km\,s$^{-1}$ in $^{12}$CO($J$=2--1). Figure \ref{fig2} shows the distribution of the outflow wings. One of them corresponds to N159W-N and the other N159W-S. The N159W-S outflow has red-shifted and blue-shifted lobes which show offsets of 0.1--0.15\,pc from the peak of the continuum emission. The outflow axis is along the east-west direction. The N159W-N outflow has only the blue-shifted lobe which shows an offset of 0.2 pc from the $^{13}$CO peak. It is possible that the complicated gas distribution around N159W-N may mask the possible red lobe. The size of the red-shifted and blue-shifted lobes is less than the beam size 0.2\,pc $\times$ 0.3\,pc and the upper-limit timescale of the outflow is roughly estimated to be 10$^{4}$ yrs. This is the first discovery of extragalactic outflows associated with a single protostar.  The positions of outflows in N159W-N and N159W-S coincide with YSOs identified based on the {\it Spitzer} data: 053937.56-694525.4 (hereafter YSO-N; Chen et al. 2010) and 053941.89-694612.0 (hereafter YSO-S; \citealt{Chen2010}; P2 in \citealt{Jones2005}), respectively. 

\subsection{YSO characteristics \label{s.r.yso}}
Two YSOs associated with outflows, YSO-N and YSO-S, have been studied extensively at near- to far-infrared, submillimeter, and radio wavelengths (\citealt{Carlson2012} and references therein; \citealt{Seale2014}; \citealt{Indebetouw2004}).  Using the \citet{Robitaille2006,Robitaille2007} YSO model grid and spectral energy distribution (SED) fitter, we model all the available data including the {\it Spitzer} and {\it Herschel} fluxes (1.2--500 $\mu$m),  as well as photometry we extracted from {\it Spitzer}/IRS spectra  (5--37\,$\mu$m; \citealt{Seale2009}), and the fit of the SEDs indicates that both YSO-N and YSO-S are Stage 0/I YSOs. 
The mass and luminosity are estimated to be $31\pm8$\,$M_\odot$ and $(1.4\pm0.4)\times10^5$\,$L_\odot$ for YSO-N, and $37\pm2$\,$M_\odot$ and $(2.0\pm0.3)\times10^5$\,$L_\odot$ for YSO-S. These results are consistent with those from Chen et al. (2010), who also used Robitaille fitter but without the Herschel constraints. 
The dynamical ages of the two outflows are consistent with the ages output from the SED fitter  \citep{Robitaille2006}, $\sim$10$^{4}$ yrs.

According to 3\,cm radio continuum measurements YSO-N is determined to be an O5.5V star, whereas YSO-S is not detected \citep{Indebetouw2004}, suggesting that YSO-S is in an earlier evolutionary state than YSO-N.
This is consistent with a non-detection of the He 2.113\,$\mu$m and with the weak Br$\gamma$ in YSO-S \citep{Testor2006}.
The \citet{Testor2006} near-IR VLT data revealed that YSO-S consists of at least two sources, whereas their detailed physical properties and relation with the mid-/far-infrared source are yet unknown.

\subsection{Filamentary collision in N159W-S \label{s.r.col}}

In Figure \ref{fig1}, N159W-N shows complicated $^{13}$CO distribution, whereas the source N159W-S shows relatively simple $^{13}$CO morphology. The $^{13}$CO distribution in N159W-N consists of several filaments which are elongated generally in the direction from the northeast to southwest, and N159W-S is located at the tip of a V-shaped distribution of two filaments. We shall focus on N159W-S in the following to describe the filament distribution and the high-mass young star, because the simple morphology allows us to understand the physical process unambiguously. 

Figure \ref{fig3} shows the two filaments toward N159W-S (Figure \ref{fig3}(a) the whole velocity range, Figure \ref{fig3}(b) red-shifted filament, and Figure \ref{fig3}(c) blue-shifted filament, respectively). The two filaments overlap toward N159W-S, where the $^{13}$CO intensity and linewidth are significantly enhanced. Figures \ref{fig3}(d-h) show position-velocity diagrams taken along the two filaments. We see that the filaments have small velocity span of 3\,km\,s$^{-1}$ in the north of N159W-S which shows significantly enhanced velocity span of 8\,km\,s$^{-1}$ at the 15\,\% level of the $^{13}$CO peak in Figure\,\ref{fig3}(g). 
An HST image at near infrared indicates that the red-shifted filament is extended toward the south beyond N159W-S (Carlson et al. 2015 in preparation), while no CO emission is detected there in our $^{12}$CO or $^{13}$CO observations with ALMA. We also find that the blue-shifted filament has its extension beyond N159W-S in $^{13}$CO. So, although the filaments are apparently terminated toward N159W-S, they are actually more extended, placing N159W-S in the intersection of the two filaments. 

In N159W-S the longer red-shifted filament in the east is highly elongated and mildly curved, having a length of 10\,pc, while the other blue-shifted filament in the west is straight and elongated by 5\,pc. N159W-S clearly demonstrates that a high-mass YSO with bipolar outflow is formed toward the intersection between the two thin filaments, and the velocity dispersion is significantly enhanced in the intersection. 

Based on these results we set up a hypothesis that formation of N159W-S was triggered by the collision between the two filaments. We first describe a possible scenario for N159W-S and then discuss the observational constraints on the collision and high-mass star formation. The two crossing filaments overlapping toward N159W-S give direct support for the present scenario. The lower limit for the relative velocity in the collision is given by the velocity difference of the two filaments 2--3\,km\,s$^{-1}$. The actual collision velocity should be higher than 2--3 km s$^{-1}$ because of the projection effect. According to the magneto-hydro-dynamical numerical simulations of two colliding molecular flows by \citet{Inoue2013}, the collision-shocked layer enhances isotropic turbulence, independent of the direction of the collision, and the velocity span in the shocked layer is similar to the relative collision velocity. The simulations by \citet{Inoue2013} for a velocity difference of 20 km s$^{-1}$ allow us to scale the relative velocity to $\sim$10 km s$^{-1}$ with basically the same physical process. We therefore assume the velocity span in N 159W-S, 8 km s$^{-1}$, as the actual collision velocity. This implies the relative motion of the two filaments is nearly vertical, roughly 70$^\circ$, to the line of sight.

\section{Discussion on the high-mass star formation processes \label{s.discussion}}

Since the rest of the filaments show no sign of velocity dispersion enhancement with high-mass star formation, we assume that the non-interacting filaments retain the initial condition prior to the collision. 
The line-mass, mass per unit length, in the filaments changes from region to region by an order of magnitude. In order to estimate the typical mass of the filaments associated with N159W-S clump for the following discussion, we pick up two segments of 1.5\,pc and 1.8\,pc in length and 0.7\,pc and 0.6\,pc in full width at a 35\,\% level of the $^{13}$CO peak for the red-shifted and blue-shifted filaments, respectively, as indicated in Figures 3(b) and 3(c). 
{Below the 35\,\% level, it is hard to estimate the line-mass of the individual filaments separately due to overlapping. Above this level, the mass sampled becomes underestimated.
We estimate the total mass of these two segments to be $2.9\times10^3$\,$M_\odot$ from $^{12}$CO($J$=2--1) for a velocity range 234\,--\,240\,km\,s$^{-1}$. 
We then estimate the average line-mass of these two filaments to be $8.9\times10^2$\,$M_\odot$\,pc$^{-1}$. 
The filaments are not detected in the Band 6 continuum at the 3\,$\sigma$ noise level of the molecular mass density $1.6\times10^3$\,$M_\odot$\,pc$^{-1}$, which is higher than the above CO-based line-mass density of the filaments.
 
This suggests that the collision took place in a timescale, $\sim$0.5\,pc divided by 8\,km\,s$^{-1}$, i.e., $\sim$6 $\times$ 10$^{4}$ years ago. We assume that formation of the high-mass star initiated at the same time. By using the stellar mass 37 $M_{\odot}$, the average mass accretion timescale of the star formation is given as 37\,$M_{\odot}$/$6 \times 10^{4}$\,yrs $\sim$6 $\times$ 10$^{-4}$ $M_{\odot}$ yr$^{-1}$. This rate is well in accord with the theoretical estimate around 10$^{-3}$ $M_{\odot}$ yr$^{-1}$ and satisfies the criterion to form high-mass stars by overcoming the stellar feedback \citep[e.g.,][]{Wolfire1986}. The small outflow timescale 10$^{4}$\,yrs is consistent with this picture involving rapid high-mass star formation.

The present case of N159W-S has shown that the high-mass star having 37\,$M_{\odot}$ is formed in a turbulent condition created by the collisional shock. 
The mass of the N159W-S clump is estimated to be $4\times10^3$\,$M_\odot$ toward its CO peak. 
There is no sign of such dense clumps over the rest of the filament according to our ALMA data, either in CS($J$=2--1) data, whose line-mass detection limit is about 150\,$M_\odot$\,pc$^{-1}$, or in dust emission data, whose line-mass detection limit is about $1.6\times10^3$\,$M_\odot$\,pc$^{-1}$ by assuming a filament width of 0.6\,pc. 
This offers an interesting possibility that high-mass stars do not necessarily require dense cloud cores as the initial condition. Instead, high velocity colliding molecular flows are able to efficiently collect mass into a cloud core non-gravitationally. \citet{Inoue2013} discuss that the mass flow in the collision can be efficiently converged into a shock-induced core due to the oblique shock effect, and that self-gravity is not important in the beginning of the high-mass star formation, while soon later, in the shock-collected core, self gravity will play a role to form the stellar core \citep[see also][]{Vaidya2013}. 

In the Milky Way we see increasing observational evidence for cloud-cloud collisions which trigger high-mass star formation. Four super star clusters, Westerlund2, NGC3603, RCW38 and DSB[2003]179, are found to be formed by collisions between two clouds \citep{Furukawa2009,Ohama2010,Fukui2014,Fukui2015}. Isolated O stars with H$\,${\sc ii} region, M20, RCW120 etc., are also suggested to be triggered by cloud-cloud collisions \citep{Torii2011, Torii2015}.  N159W-S is in the very early stage of star formation as indicated by the non-detection of ionized gas, as well as by the collision scenario and SED models. Therefore N159W-S is an optimal source to study filamentary collision leading to star formation. 
It has been shown that the youngest O stars are formed coevally in a duration of $\lesssim10^{5}$\,yrs in NGC3603 and Westerlund1 by careful measurements of stellar ages with HST and VLT \citep{Kudryavtseva2012}. We have here an independent estimate of the stellar age by taking the advantage of the simple cloud morphology in N159W-S, and the present time-scale estimate is consistent with that of \citet{Kudryavtseva2012}.

\section{Conclusions \label{s.conclusion}}

In this Letter we presented the $^{13}$CO ($J$ = 2--1) observations with ALMA on the active star-forming region N159 West in the LMC. We have found the first two extragalactic protostellar molecular outflows toward young high-mass stars, whose dynamical timescale is $\sim$10$^{4}$ yrs. One of the two stars N159W-S is clearly located toward the intersection of two filamentary clouds. We set up a hypothesis that two filaments collided with each other $\sim$10$^{5}$ yrs ago and triggered the formation of the high-mass star. The results demonstrate the unprecedented power of ALMA to resolve extragalactic star formation.

\acknowledgments

The authors thank the anonymous referee for his/her helpful comments. This paper makes use of the following ALMA data: ADS/JAO.ALMA\#2012.1.00554.S. ALMA is a partnership of ESO (representing its member states), NSF (USA) and NINS (Japan), together with NRC (Canada), NSC and ASIAA (Taiwan), and KASI (Republic of Korea), in cooperation with the Republic of Chile. The Joint ALMA Observatory is operated by ESO, AUI/NRAO and NAOJ. This work was supported by JSPS KAKENHI grant numbers 22244014, 22540250, 22740127, 23403001, 24224005, 26247026; by JSPS and by the Mitsubishi Foundation.  MM and ON are grateful for support from NSF grant \#1312902.

\clearpage

\begin{figure}
\epsscale{.99}
\plotone{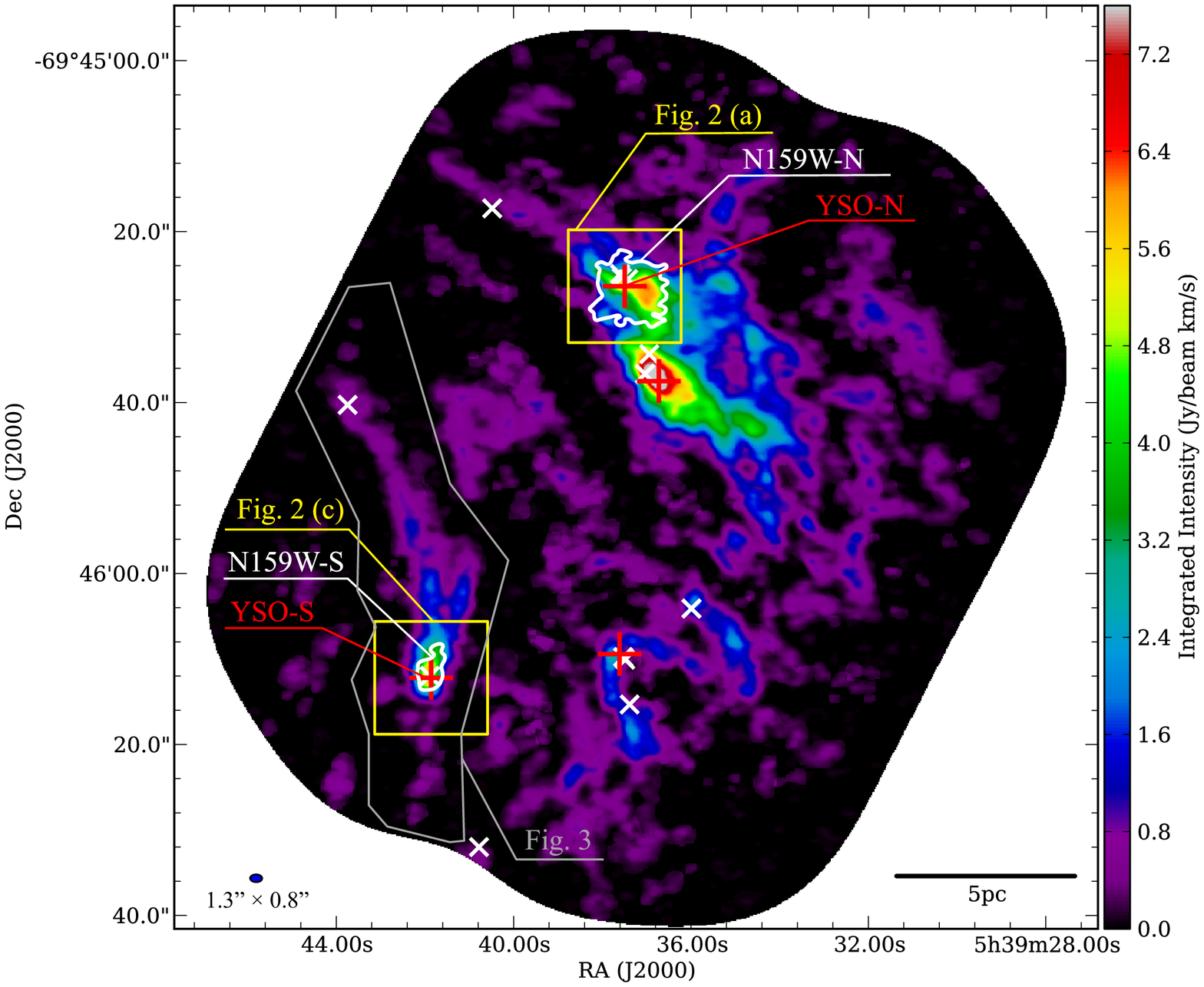}
\caption{Velocity-integrated intensities of $^{13}$CO($J$=2--1) toward N159W with the 12-m array observation is presented. White and red crosses show the YSOs (Chen et al. 2010) and the Band 6 continuum peaks of the ALMA observation, respectively. White contours are plotted at the 5\,$\sigma$ level of the Band 6 continuum emission. 
\label{fig1}}
\end{figure}

 \begin{figure}
\epsscale{.80}
\plotone{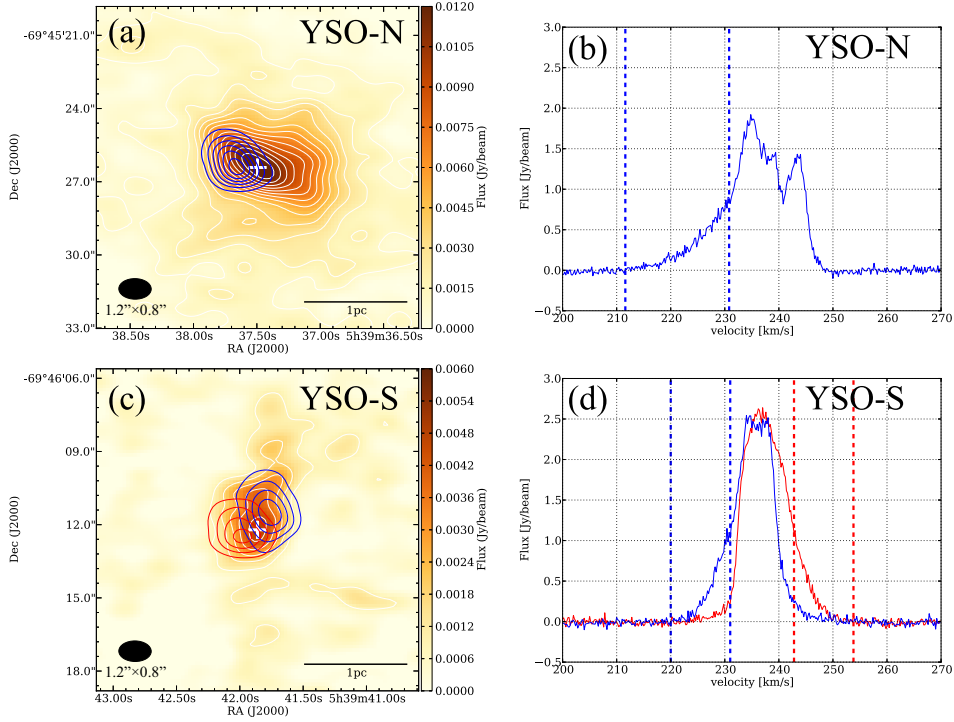}
\caption{Distributions of high-velocity wings from two outflows: (a) and (b) for YSO-N and (c) and (d) for YSO-S.
(a, c) Image and white contours show the Band 6 continuum emission.  Contours are plotted every 0.9\,Jy\,beam$^{-1}$. 
Red and blue contours show images of $^{12}$CO($J$=2--1) integrated over the velocity ranges shown in red and blue dashed lines in (b) and (d). Contours are plotted every 1.1\,Jy\,beam$^{-1}$\,km\,s$^{-1}$. White cross depicts the Band 6 continuum peak. (b, d) The red and blue lines show averaged spectra of the outflow wing over the regions inside the red and blue contours, respectively, in (a) and (c).
\label{fig2}}
\end{figure}

\begin{figure}
\epsscale{1.}
\plotone{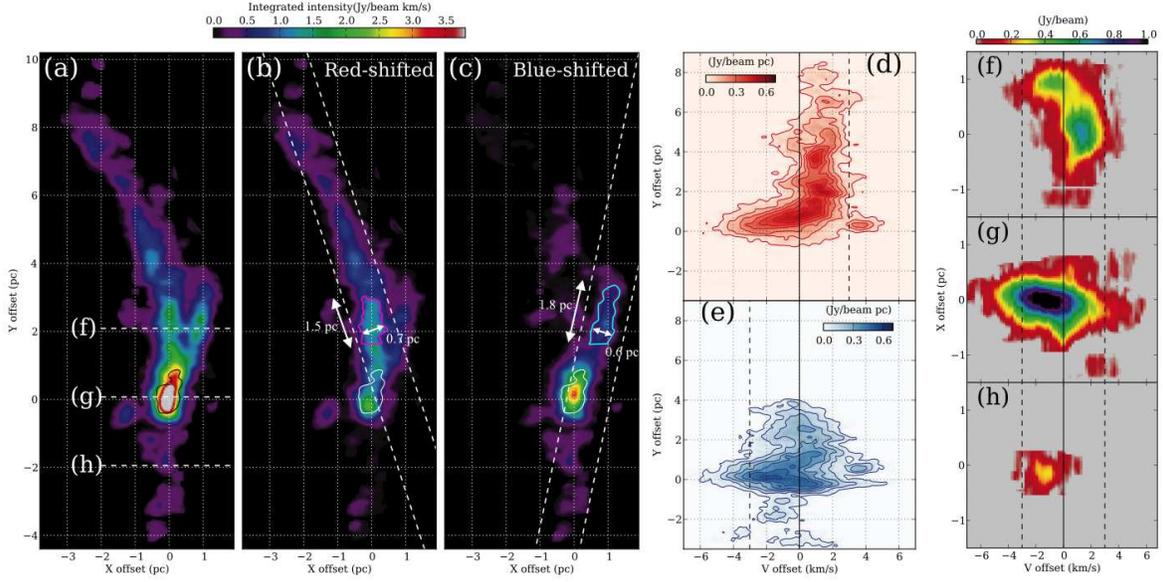}
\caption{Spatial and velocity distributions of two filaments in N159W-S. (a) $^{13}$CO($J$=2--1) intensity map integrated over a velocity range 234\,--\,240\,km\,s$^{-1}$. The X offset in R.A. (J2000.0) and Y offset in Dec. (J2000.0) are from (4$^{\rm h}$39$^{\rm m}$41\fs88, -69\arcdeg46\arcmin12\farcs2). (b, c) The $^{13}$CO($J$=2--1) map integrated over a velocity range between the dashed lines of (d) and (e), respectively. Magenta and cyan lines show the regions used for mass estimate in Section 4 for the velocity range same with that in (a). The arrows depict the lengths and widths of the regions. The black/white contours plotted in (a)--(c) indicate the 5\,$\sigma$ level of the Band 6 continuum emission. (d, e) $^{13}$CO($J$=2--1) position-velocity diagrams along the region between the dashed lines in (b) and (c), respectively, where the central velocity is 237\,km\,s$^{-1}$. Contours are plotted at every 6\,$\sigma$ from 3\,$\sigma$ ($1\,\sigma = 0.013$\,Jy\,beam$^{-1}$\,pc). (f--h) Position-velocity diagrams of $^{13}$CO($J$=2--1) along the dashed lines in (a).
\label{fig3}}
\end{figure}

\clearpage

\end{document}